\documentclass[%
groupedaddress,
preprint,
showpacs,preprintnumbers,
 amsmath,amssymb,
 aps,
prb,
]{revtex4-1}
\usepackage{graphicx,epsfig}
\usepackage{dcolumn}
\usepackage{bm}
\usepackage[colorlinks=false,linkcolor=blue]{hyperref}
\begin{document}


\title{Phonon anomalies and lattice dynamics in superconducting oxychlorides Ca$_{2-x}$CuO$_2$Cl$_2$}

\author{Matteo d'Astuto}
\email{matteo.dastuto@impmc.upmc.fr}

\author{Ikuya Yamada}
\altaffiliation[Present address: ]{Nanoscience and Nanotechnology Research Center (N2RC), Research Institutes for the Twenty-First Century - Osaka Prefecture University
1-2 Gakuen-cho, Naka-ku, Sakai, Osaka 599-8570, Japan} 

\author{Paola Giura} 
\author{Lorenzo Paulatto}
\author{Andrea Gauzzi}

\affiliation{Institut de Min\'eralogie et de Physique des
Milieux Condens\'es (IMPMC), UMR CNRS 7590, Universit\'e Pierre et Marie Curie - case 115, 4, place Jussieu, 75252 Paris cedex 05, France}
\author{Moritz Hoesch} 
\altaffiliation[Present address: ]{Diamond Light Source, Didcot OX11 0DE, United Kingdom}
\author{Michael Krisch}
\affiliation{European Synchrotron Radiation Facility,
BP 220, F-38043 Grenoble Cedex, France}

\author{Masaki Azuma}
\author{Mikio Takano} 
\affiliation{Laboratory of Solid State Chemistry, Institute for Chemical Research, Kyoto University, Uji, Kyoto-fu, 611-0011 JAPAN}

\begin{abstract}
We present a comprehensive study of the phonon dispersion in an underdoped, superconducting Ca$_{2-x}$CuO$_2$Cl$_2$ crystal. We interpret the results using lattice dynamical calculations based on a shell model, and we compare the results, to other hole-doped cuprates, in particular to the ones isomorphic to La$_{2-x}$Sr$_x$CuO$_4$ (LSCO).
We found that an anomalous dip in the Cu-O bond stretching dispersion develops in oxychlorides with a simultaneous marked broadening of the mode. The broadening is maximum at $\approx (\pi / (2a) ~ 0 ~ 0)$ that corresponds to the charge-modulations propagation vector.
Our analysis also suggests that screening effects in calculations may cause an apparent cosine-shaped bending of the Cu-O bond-stretching dispersion along both the ($q$ 0 0) and ($q$ $q$ 0) directions, that is not observed on the data close to optimal doping. This observation suggests that the discrepancy between experimental data and \textit{ab-initio} calculations  on this mode originates from an overestimation of the doping effects on the mode. 
\end{abstract}

\pacs{74.72.Gh, 63.20.D-, 63.20.kd, 78.70.Ck}

\keywords{Superconductivity, Phonons, Hole-doped Cuprate, Phonon-electron interactions, Phonon dispersion, Inelastic X-ray Scattering} 

\maketitle

\section{\label{intro}Introduction}

The discovery of high temperature superconductivity (HTS) in cuprates was originally motivated by the search for superconductors with strong electron-phonon coupling\cite{bm}. The first HTS compounds, La$_{2-x}$Ba$_x$CuO$_4$ (LBCO) and LSCO, do indeed display a very large isotopic effect for oxygen, although the doping dependence of this effect turns out to be very anomalous\cite{Crawford-isoLaCuO}. 
Later studies pointed out that electron-phonon coupling may not be the main origin of the Cooper pairing\cite{nat-phys-bonn-rv}, in particular for optimal doping. But its role across the phase diagram of cuprates is still controversial, in particular in its under-doped region. An important point is that single electron band approximations, such as Density Functional Theory (DFT) find a small electron-phonon coupling \cite{bohnen,giustino}, but they possibly miss many body effects, as it was suggested that large couplings may exist in the presence of strong electron-electron correlations \cite{rosch-gunn}. 

An important point stressed in DFT based investigations on electron-phonon coupling in cuprates is that these calculations seem to reproduce well the marked softening with a cosine dispersion, that develops under doping for longitudinal Cu-O bond-stretching mode along the direction ($q$ 0 0) of such bonds (modes with $\Delta_1$ character). 
At the same time, large discrepancies between those calculations for the same mode, but along the diagonal direction ($q$ $q$ 0) (modes with $\Sigma_1$ character) are found\cite{falter-lsco,bohnen}. It has been suggested that the physical origin of the anomaly in the dispersion of the Cu-O bond-stretching modes is different for the two directions \cite{bohnen,giustino}.

Underdoped cuprates exhibit a static, or quasi-static, modulation of the charge, characterized by a  stripe pattern parallel to the Cu-O bonds, which can give clear diffraction peaks both in the charge \cite{ tranquada1,abbamonte} and spin channel, for doping close to $\delta \sim 1/8$ with a propagation vector of $\approx \pi / (2a)$, \textit{i.e.} a periodicity of $\approx 4a$, where $a$ is the lattice parameter along the Cu-O bonds. 
Besides the above mentioned Cu-O bond stretching anomaly with cosine shape, an additional marked dip and a sizable broadening of the line-shape is observed for the same phonon branch, close to the wave stripe vector  $\mathbf{q}=(0.25, 0, 0)$ \cite{mcqueeney,pinbrief,reznik,dastuto-laba,reznik-rev}; it has been established that such feature can not be simulated by DFT \cite{reznik-gunn}. 

Large electron-phonon coupling can drive both phonon softening and broadening as it as been established, for example, in the case of classical superconductors such as, \textit{e.g.} Nb \cite{butler} and, more recently, in ${\mathrm{YNi}}_{2}{\mathrm{B}}_{2}\mathrm{C}$ \cite{weber-YNi2B2C}.
For cuprates the issue remains controversial. In order to elucidate this aspect, we investigate the lattice dynamics in superconducting, under-doped Ca$_{2-x}$CuO$_2$Cl$_2$, using Inelastic X-ray Scattering (IXS) and compare our results with lattice dynamics calculations based on a shell model. 

\begin{figure}[t]
\includegraphics[width=\linewidth]{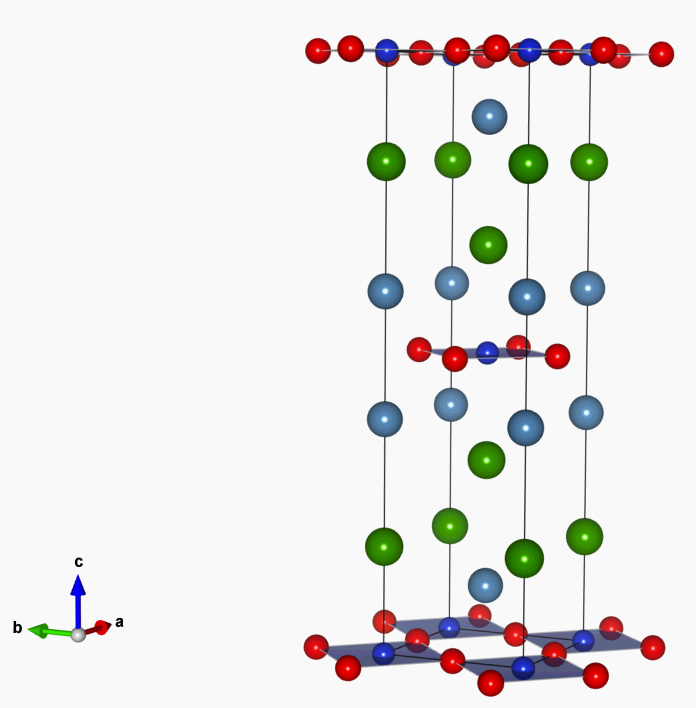}
	 \caption{\label{ccoc structure} (Color on line) Drawing\cite{VESTA} of the Unit cell of Ca$_2$CuO$_2$Cl$_2$ \cite{yamada-ccoc2}. The square coordination of copper with its four nearest-neighborhood oxygens ions in the CuO$_2$ planes is shown. The chlorine ions are located in the apical site above the copper.}
 \end{figure}

Doping Ca$_2$CuO$_2$Cl$_2$ oxychloride systems leads to a new family of HTS superconductors with Tc up to 43 K\cite{hiroi,yamada-ccoc}.  Their layered structure is similar to the cuprate one,  characterized by CuO$_2$ plane, but without oxygen in the blocks separating these planes (see Fig. \ref{ccoc structure}, left panel). 
These compounds represent an ideal test bench for the physics governing
the high Tc superconductivity that is commonly believed to depend only
on the CuO$_2$ planes, independently from the details of the oxide 
blocks between them. 
In particular, oxychlorides have a chlorine ion that replaces the apical oxygen. The apical oxygen is important for the dynamics of the anionic octahedra surrounding the copper cation on most of the hole-doped cuprates. 
Moreover, oxychlorides are ideal for studying phonon dispersions, particularly the one of the copper-oxygen bond stretching mode by means of IXS spectroscopy, because they are composed only of low Z, thus yielding a higher IXS signal, due to the weaker photoelectric absorption compared to the cuprates, 
and also thanks to their simple 1-layer structure with small disorder, which reduces the number of modes to be measured and calculated. We note that these characteristics should also facilitate \textit{ab-initio} calculations.
Underdoped oxychlorides show a modulation in the charge density, as their cuprate counterpart. However, this modulation is observed only in real space, using Scanning Tunelling Microscopy (STM) \cite{hanaguri}, with a periodicity of $\approx 4a$  as for the stripes, but, up to now, not in diffraction. 

In this work, we present a study of the phonon dispersion in superconducting Ca$_{2-x}$CuO$_2$Cl$_2$ doped with Ca vacancies. The study is as complete as possible, considering limitation from the crystal mosaic for the particular technique used (IXS), and the limited beam-time available on such highly specialized synchrotron experiment. We interpret the results using lattice dynamical calculations based on a shell model, and we compare our results, to those obtained for other hole-doped cuprates, in particular the ones isomorphic to LSCO (T-structure)\cite{pintrev1,reznik-rev,tajima-IR-cuprates}.
We found that an anomalous dip in the Cu-O bond stretching dispersion develops also in oxychlorides 
with a simultaneous marked broadening of the mode. The maximum of the broadening appear at the charge modulation propagation vector of $\approx (\pi / (2a) ~ 0 ~ 0)$. 
In addition to this effect, by comparing our results on the longitudinal Cu-O bond-stretching mode in superconducting Ca$_{2-x}$CuO$_2$Cl$_2$ to our model, we found that screening effect may lead to a simultaneous bending down of the whole dispersion along both the ($q$ 0 0) and ($q$ $q$ 0) directions. 

\section{\label{sec:methods} Methods}

\subsection{Crystal growth and characteristics}

\begin{figure}[t]
\includegraphics[scale=0.35]{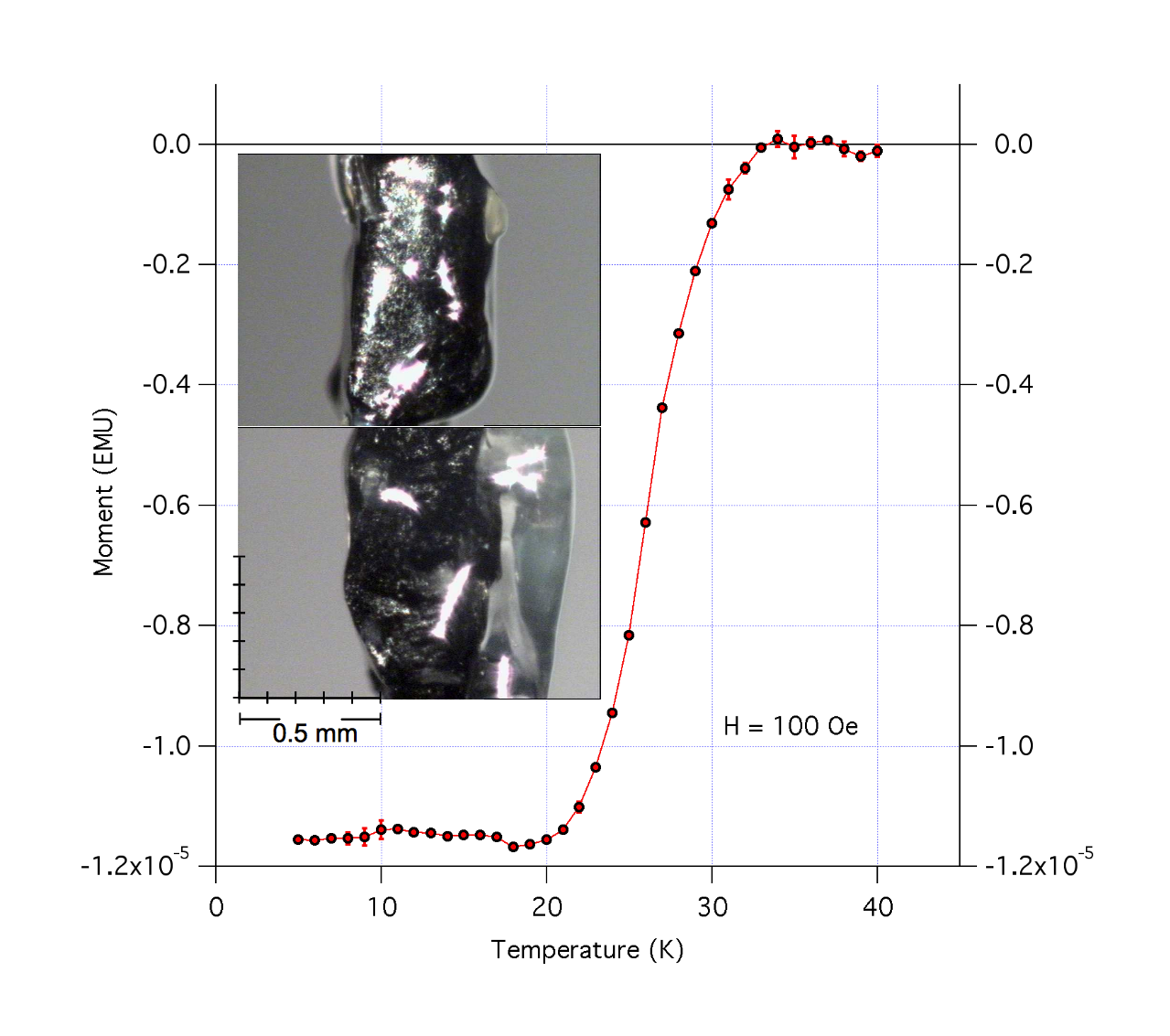} 
	 \caption{\label{ccoc-op2b1-MssnrP} (Color on line) Chart\cite{igor} of the magnetization \textit{vs} temperature across the superconducting transition at 100 Oe for the sample 2b measured by IXS. Inset, front and side view of the crystal, as measured in diffraction and IXS, in a grease film, with the glass fiber holder. }
 \end{figure}

We studied three different crystals of Ca$_{2-x}$CuO$_2$Cl$_2$, selected from several ones, of sizes on the order of 0.5 to 1 mm, from different batches, after a screening of their superconducting and crystalline properties as measured by the Mei\ss ner effect in a SQUID magnetometer (Quantum design MPMS\copyright ) and the diffraction on a four-circle diffractometer equipped with a Mo K$\alpha$ anode, a kappa-type goniometer and a CCD detector (Oxford diffraction Xcalibur\copyright). 
We selected two of the crystals with nominal Ca vacancies doping x=0.2, grown at high-temperature and high-pressure of $1250^{\circ}$C, 5.5 GPa\cite{yamada-ccoc2,yamada-ccoc}, and a parent insulating Ca$_{2}$CuO$_2$Cl$_2$ with stoichiometric composition. 
The best one, sample 2b, used for all the measurements of \textit{in-plane} dispersion, of about 0.1 mm$^3$ volume (see Fig. \ref{ccoc-op2b1-MssnrP}), shows a superconducting transition temperature onset at T$_c^{onset}$=33 K, with a well defined diamagnetic jump on a temperature window $\Delta T \approx$ 8 K from 10 to 90\% of the magnetization, as shown in Fig. \ref{ccoc-op2b1-MssnrP}. This corresponds to an underdoped compounds with a
 Ca deficiency x$\sim$0.16$\pm$0.01, slightly lower than the nominal one. 
The sample was protected from moist by a film of grease (Apizon N\copyright), in order to avoid formation of hydroxides, during all the measurements. 
The grease film was also used to hold it on the borosilicate glass sample holder for diffraction and Inelastic X-ray Scattering (IXS), as shown in Fig. \ref{ccoc-op2b1-MssnrP} (inset).
The second doped crystal, named 1b, has also a nominal vacancy doping x=0.2, and similar crystalline and superconducting properties.

\subsection{Inelastic X-ray Scattering}
 
The sample 2b was mounted in a vacuum chamber with Kapton\copyright 
~windows, with the scattering plane ($(1,1,0)$,$(0,0,1)$) for
measurements along $(1,1,0)$ and ($(1,0,0)$,$(0,1,0)$) for measurements along $(1,0,0)$. 
The angular width of the (200) Bragg reflection rocking curve was 
$0.01^{\circ}$ FWHM, proving excellent single crystal quality \textit{in-plane}.  
The second sample, 1b, was mounted with the direction $(0,0,1)$ in the plane of scattering, but it revealed a large mosaic in the c* direction, preventing us from completing the measurements along $(0,0,1)$. This also prevented us for doing measurements on modes with \textit{in-plane} propagation but \textit{out-of-plane} polarization as, notably, the oxygen buckling modes. 
The third, undoped, sample showed a mosaic too large for doing any phonon measurement.
Phonons were measured  from (1,1,0) to (2.2, 2.2, 0) along $(1,1,0)$ and from (2,0,0) to (3,0,0) along $(1,0,0)$ in longitudinal geometry, and from (3,0,0) up to (3, 0.5, 0) in transverse one.  

The main monochromator was set to the Si (9 9 9) Bragg reflection, with a
wave-length of 0.6968 \AA (17794 eV) and an energy resolution
$\Delta$E = 3.0 $\pm$ 0.2 meV (see Ref. \onlinecite{krisch-lss,verbeni,verbeni-rsi} for details).
 The back-scattered beam is focused by a 
platinum-coated toroidal mirror,
which provided a focal spot at the sample position of 0.270
(horizontal) and 0.080 (vertical) mm$^2$ FWHM at the sample position. 
The IXS spectrometer \cite{masciovecchio1} was used 
with a set-up allowing simultaneous measurements from 5 analyzers.  
Analyzer slits were open to (60 $\times$ 60) mm, thus integrating over
a solid angle of 0.6$^{\circ}$, amounting to a Q
resolution of $\pm$0.416 nm$^{-1}$ corresponding to approximately 0.06 of
the Brillouin Zone along $(1,0,0)$. 

\begin{figure}
	 \includegraphics[width=\linewidth]{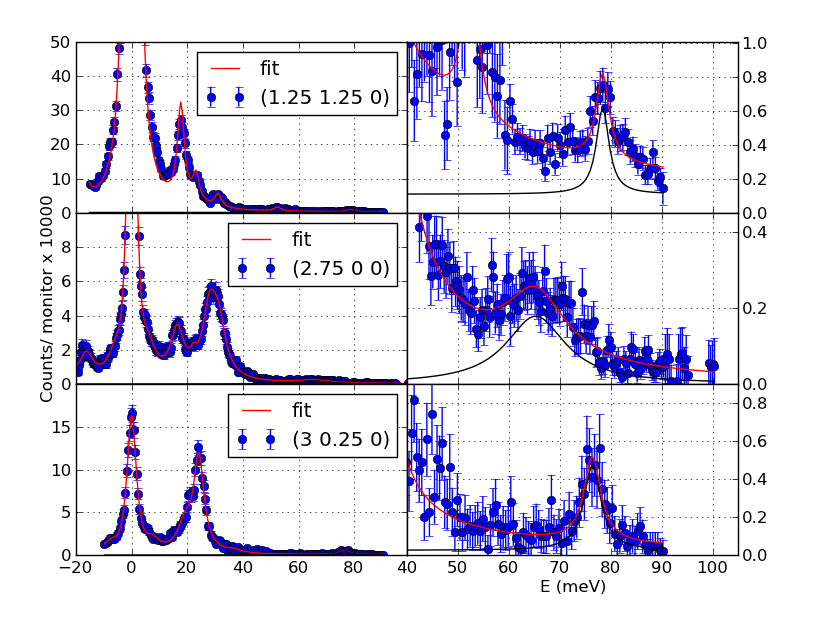}
	  \caption{\label{example-raw} (Color online) Plot\cite{Matplotlib} of the measured IXS intensities as a function of energy on Ca$_{2-x}$CuO$_2$Cl$_2$ for the propagation vector corresponding to the middle of the Brillouin Zone along three different symmetry \textit{in-plane} lines. Left column shows the overall data, while the right column are the respective zoom at the high energy of the Cu-O bond stretching mode. 
Top panel: Longitudinal  at (0.25 0.25 0), with $\Sigma_1$ symmetry.   
Middle panel: Longitudinal  at (0.25 0 0), with $\Delta_1$ symmetry.   
Bottom: Transverse at (0 0.25 0), with $\Delta_3$ symmetry.   
Red lines are fit to the data described in the text. The contribution from the Cu-O bond stretching mode are singled out as a black line. 
	  }
 \end{figure}

Fig. \ref{example-raw} shows representative IXS spectra for three different directions, longitudinal  along ($q$ $q$ 0), with $\Sigma_1$ character, longitudinal along the ($q$ 0 0) direction, with $\Delta_1$ character and transverse along the equivalent (0 $q$ 0) direction, with $\Delta_3$ character, for the same value $q$=0.25. 
Continuous lines are a model, consisting of Lorentzian function, convoluted with the instrumental function. The model is chosen to fit the data using a $\chi^2$ minimization procedure \cite{james}. 

In Ca$_{2-x}$CuO$_2$Cl$_2$ the contrast between the elastic line and the low energy mode on one side and high energy mode on the other, it is not as strong as in other cuprates. The ratio is determined by the ions number of electrons, as the the IXS scattering yield is proportional to $Z^2$ and the photoelectron absorption to $Z^4$, so that the high $Z$ elements dominate in a system with different atomic species, while the high energy mode come almost exclusively from light element vibrations\cite{dastuto-jdn}. In other cuprate systems, measuring the high energy mode dispersion is therefore possible only at few ($\sim$ 10) Kelvin, where the Bose factor is strongly reduced for modes below 25 meV, compared to the intensity at room temperature\cite{dastuto-ncco-long}. 
The measurements presented here are, in turn, all taken at room temperature, so that the contrast in the intensities between high and low energy is largely due to the Bose contributions, still allowing the measurements of all high energy phonon modes. 

\subsection{Calculations}

\begin{figure}
	 \includegraphics[width=\linewidth]{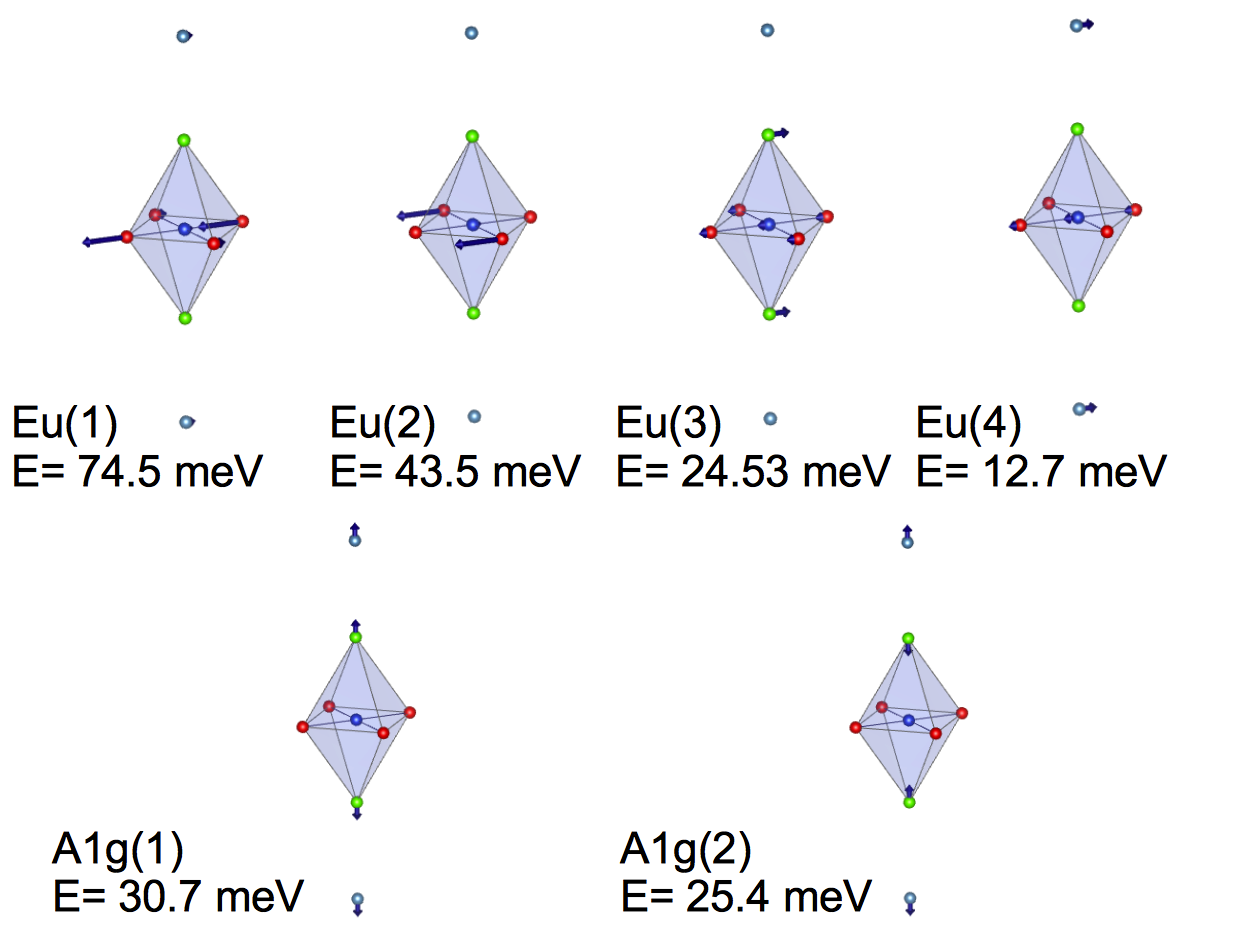}
	  \caption{\label{Gamma-EV} (Color online) Drawing\cite{VESTA} of the atomic displacements of the \textit{in-plane} modes we have measured at  the zone center $\Gamma$, as calculated by our lattice dynamical model described in the text. For each mode, we give its character, and the calculated frequency at $\Gamma$. We highlight the octahedron formed by the anions, with the two apical chlorines and the four square coordinated oxygens, around the copper, in order to single out the differences and similarities with the analogue displacements in hole-doped cuprates with T structure, as \textit{e.g.} shown in Ref. \onlinecite{tajima-IR-cuprates}. The calcium cations lies outside the octahedron, on the vertical of the apical chlorine one. }
 \end{figure}

Phonon dispersion are simulated using the lattice dynamical calculation 
package OpenPhonon\cite{mirone}, based on 
a shell model, further adapted for compilation with python 2.7 \cite{MPD}. 
In order to be consistent with previous studies on the phonon dispersion in cuprates 
by INS \cite{pintrev1,reznik-rev} and IXS \cite{dastuto-ncco, dastuto-laba,dastuto-hg1201}
we used a common potential model for cuprates, 
where the interatomic potentials have been derived from a comparison  
of the INS results for different  HTcS compounds by Chaplot {\it et al.} \cite{chaplot}.
To this model a screening of the Coulomb potential is added, 
in order to simulate the effect of the free carriers introduced by doping. 
In the same way as described by Chaplot (see Ref. \onlinecite{chaplot})  
for metallic $\mathrm{La_{2-x}Sr_xCuO_4}$ and $\mathrm{YBa_2Cu_3O_7}$, we replaced the 
long-range Coulomb potential $V_c(q)$ by $V_c(q)/\epsilon(q)$, and for 
the dielectric function we took the semi-classical 
\textit{Thomas-Fermi} limit of $\epsilon(q)=1+\kappa_s^2/q^2$,
where $\kappa_s^2$ indicates the screening vector. In our calculation, we estimate a screening wave number $\kappa_s = \frac{4}{a} \sim 0.4$ \AA $^{-1}$, where $a$ is the lattice parameter, which 
corresponds to a doping of about 10 \%  according to Ref. \onlinecite{chaplot}.
For ions and bonds not included in the previous model we used the parameters used by Mittal et al. in Ref. \onlinecite{mittal-MFX} for Cl and by Chaplot in Ref. \onlinecite{chaplotYBCO} for Ca as a starting point of our refinement. 
We used the experimental lattice parameters as reported in Ref. \onlinecite{yamada-ccoc}.
The solution is stable with the set of atomic potential found, without adding any additional force constant to equilibrate the fictitious forces originating from anharmonic contribution when using real structural parameters in a quasi-harmonic approximation. 
A summary of the ions and bonds parameters are given in Table \ref{table:ions_param} and \ref{table:bond_param} respectively.
In Fig. \ref{Gamma-EV} we show the main atomic displacement pattern as calculated by our model at the zone center $\Gamma$, for the \textit{in-plane} modes we have measured. 

\begin{table}
\caption[]{Ions parameters of the lattice dynamical shell model described in the text. 
\label{table:ions_param}} 
\begin{ruledtabular}
\begin{tabular}{cccc}
$k$ & $Z(k)$ & $Y(k)$ & $K(k)$(nm$^{-1}$) \\\hline
O  & -1.56 & -3.0 & 1800\\
Cl & -0.92 & -1.0 & 1800\\
Cu & 1.64 & 3.0 & 2000\\
Ca & 1.66 & 3.0 & 700\\
\end{tabular}
\end{ruledtabular}

\caption[]{Bonds parameters of the lattice dynamical shell model described in the text. 
\label{table:bond_param}} 
\begin{ruledtabular}
\begin{tabular}{cccc}
k-k' & $A_{kk'}$ (eV) & $R_{kk'}$ (\AA) & $C_{kk'}$ (ev \AA$^6$) \\\hline
Cu-O & 4000 & 0.228 & \\
Cu-Cl & 4000 & 0.29 & \\
O-O & 2000 & 0.284 & 100 \\
Ca-O & 1600  & 0.304 & \\
Cl-Cl & 2000 & 0.34 & \\
Ca-Cl & 1600 & 0.33  & \\
\end{tabular}
\end{ruledtabular}
\end{table}

Note that, from now on, we will use the character of the mode along a particular dispersion line as a label for the line itself (\textit{e.g.} $\Delta_1$ for the line ($q$ 0 0) in longitudinal geometry) , and the zone center character as a label for the whole mode dispersion (\textit{e.g.}  Eu(1) along $\Delta_1$ for the Cu-O bond stretching mode along the line ($q$ 0 0) in longitudinal geometry). 
 
\section{\label{sec:results} Results}

IXS spectra shows overall well defined phonon modes, in general with resolution-limited energy width, with the notable exception of the Cu-O bond stretching mode (Eu(1)) along $\Delta_1$. They therefore present no difficulty for the fit procedure, in order to determine the phonon frequencies, width and intensities. 

\begin{figure*}
	\includegraphics[scale=0.65]{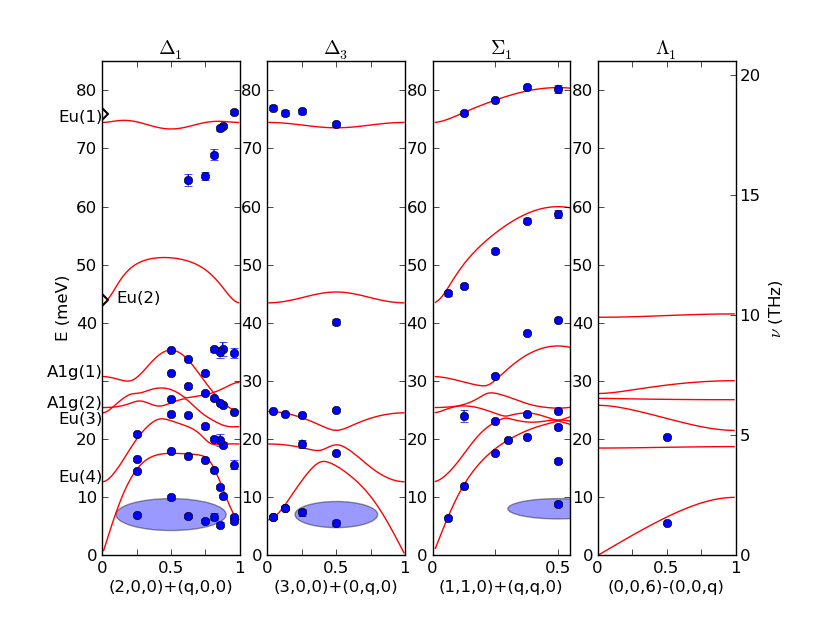}
	\caption{\label{disp} (Color online) Chart\cite{Matplotlib} of the Ca$_{2-x}$CuO$_2$Cl$_2$ phonon dispersion (blue circles) measured by IXS and calculated frequencies using the shell model of lattice dynamics described in the text (red line). We also indicate, on the left panel, the frequencies at  the zone center $\Gamma$, extrapolated from their positions in the infrared spectra reported in Ref. \onlinecite{ccoc-ir} (empty triangle-right).
	 }
\end{figure*}

\begin{figure*}	 
	 \includegraphics[scale=0.55]{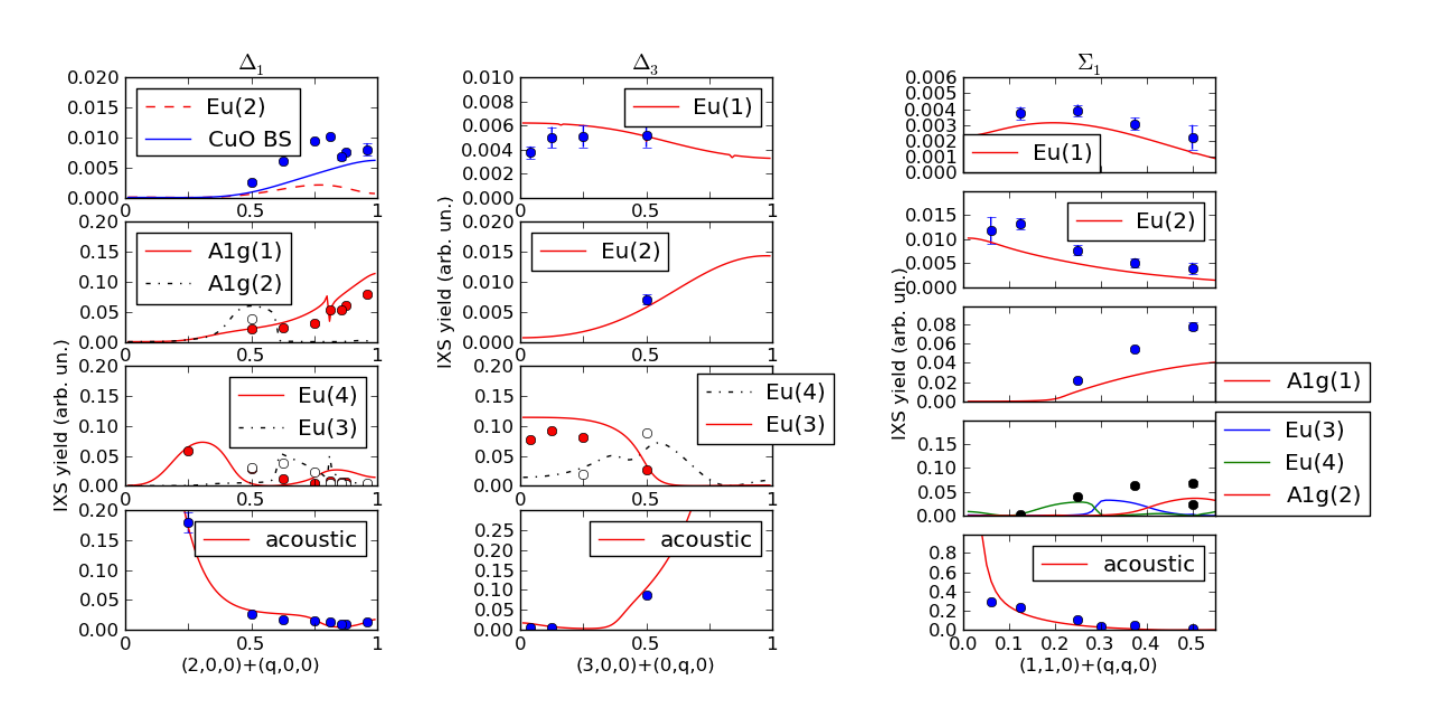}	\caption{\label{ineint}  (Color online) Calculated (lines) and experimental (symbols) inelastic scattering intensities for phonon modes in Ca$_{2-x}$CuO$_2$Cl$_2$ CuO, with character $\Delta_1$ (left panel), $\Delta_3$ (center) and $\Sigma_1$ (right panel)\cite{Matplotlib}. The intensities units are arbitrary for calculations. Experimental reported areas of peak units are in I/I$_0 \times$ meV, and are multiplied by a common factor for all modes for a given line in the Brillouin Zone. Mode assignment is based  on the closest experimental frequency to a given calculated mode. The only exception is the acoustic and(or) Eu(4) mode at q=(3, 0.5, 0), for which both assignment are proposed (and match) the calculated one. Labels corresponds to the character a the zone center as explained in the text. 
	 }
\end{figure*}

The parameters obtained from the fitting of these and several others spectra are plotted in Fig. \ref{disp}, for the frequency mode dispersions, and Fig. \ref{ineint}, for the mode intensities.
In Fig. \ref{disp} the experimental frequencies obtained by IXS are compared with the calculated dispersions. We also include experimental zone center modes estimated frequencies, obtained by infrared absorption on heavily underdoped Ca$_{2-x}$Na$_{x}$CuO$_2$Cl$_2$ (x=0, 0.3) in Ref. \onlinecite{ccoc-ir}.  
In Fig. \ref{ineint}, the IXS intensities are compared with the dynamical structure factor calculated using the eigenvector simulated in our model.  
Overall, we observe a good agreement between the measured frequencies and intensities, on one side, and the calculated ones on the other side. This with the notable exception of : 
\begin{enumerate}
\item the highest energy optic mode, the Cu-O bond stretching, with $\Delta_1$ symmetry,
\item a spurious intensity at very low frequency, at about 8 meV, with flat dispersion \textit{in-plane}, 
\item the dispersion for the Eu(3) mode, the Cu-O bond-bending, along $\Delta_1$, away from the zone center, and
\item the acoustic $\Sigma_1$ mode close to the zone boundary.
\end{enumerate}

The first anomaly is well known \cite{pintrev1,reznik-rev,dastuto-ncco, dastuto-laba,dastuto-hg1201}, and widely debated \cite{giustino,rosch-gunn}, while the other anomalies are rather new. We discuss the details of our observation and possible interpretations in the following section.

In table  \ref{table:ions_param} and \ref{table:bond_param} we give the ionic and bond parameters used for the model shown in Fig. \ref{disp} and \ref{ineint}. 

Overall, the agreement between the calculated and measured dispersion is comparable to the one found using similar atomistic models\cite{pintrev1, chaplot} or even more advanced \textit{ab-initio} approaches \cite{reznik-rev,giustino} . Moreover, we find a satisfying agreement between the calculated intensities, with the experimental ones on the IXS peak with closer frequencies of the calculated dispersion. This, not only supports the assignment we suggest for the phonons branches we have measured, but it is also a proof of the quality of the model, as correct eigenvectors are more delicate to found than frequencies.  


\section{\label{sec:discussion} Discussion}

The lattice dynamical shell model has been refined only with limited tuning of the ionic and bond parameter concerning the chlorine and calcium ions, otherwise imposing the parameters from Ref. \onlinecite{chaplot}, confirming that the common potential model is appropriate for the cuprates. 

We would like to discuss now in more details our present understanding of the discrepancies between the data and the calculations.   

\paragraph{Dispersionless intensity at about 8 meV.} We observe a low energy flat mode (point 2 above), \textit{in-plane}, mainly along the  $\Delta_1$ and $\Delta_3$, directions. Along the $\Sigma_1$ direction the fitting of the central line is improved if we add some intensity in the same frequency region, but it is not well resolved, with perhaps the exception of the zone boundary. The difference is  possibly due to the larger intensity of the elastic line in the geometry corresponding to $\Sigma_1$ scattering, which is accidental, and therefore, we can not consider this anomalous excitation anisotropic from our data. In the $\Delta_1$ and $\Delta_3$ directions it appears as a rather shallow intensity between the elastic line and the acoustic mode, but with an intensity well above the background. For a few scans only it appears with a well defined maximum. 
The origin of this mode is not understood at the moment, but also observed in closely related system, for which we are presently running experimental investigations\cite{test-llb} in order to test different hypotheses of the nature of this additional mode, including defect vibrational states and phonon modes activated by symmetry breaking. We note also that several works using Raman scattering and infrared absorption in underdoped La$_{2-x}$Sr$_x$CuO$_{4+\delta}$\cite{tassini, sugai, caprara, lucarelli, venturini} found intensity at similar energy in under-doped sample, which is attributed to modes related to the stripes charge-modulations. 

\paragraph{Acoustic $\Sigma_1$ mode close to the zone boundary.} The anomaly on the $\Sigma_1$ acoustic mode close to the zone boundary concern only one point, while the rest of the dispersion fit quite well with our simulations, both for the frequencies and the intensities. It would be nevertheless interesting to further study this mode, by complementary \textit{ab-initio} calculation, to verify our findings, as well as with more detailed measurements close to the zone boundary. 

\paragraph{Cu-O bond bending mode.} The calculated frequency of the  Cu-O bond bending mode at the zone center $\Gamma$ is very close to the experimental IR position for the corresponding Eu(2) mode, but our calculation shows an upward dispersion, which does not seems to be confirmed in our data along $\Delta_3$. Indeed, the experimental intensity along $\Delta_3$ between 30 and 40 meV, seems to lie at lower energy with respect to the calculated one. However, the Cu-O bond bending mode has a relatively low yield, along both $\Delta_1$ and $\Delta_3$ comparable to the bond-stretching one, while being at lower energy. They are  therefore measured on top of the tails of the much stronger modes at lower energy, as described above, fact that complicates their detection. A further investigation of this dispersion along $\Delta_1$ and $\Delta_3$  would be possible only at low temperature, as discussed above. Note that, on the contrary, for the corresponding Eu(2) mode along $\Sigma_1$ the mode is reasonably well resolved, and match very well both the calculated frequency dispersion and the intensity. 

\paragraph{Cu-O bond stretching mode.} Concerning the Cu-O bond stretching along the bond direction, ($q$, 0, 0), we calculate a symmetrical dispersion in the extended zone, similarly to other cuprate with T-structure. Frequencies at both zone center $q$=0 and extended zone boundary $q$=2$\frac{\pi}{a}$ are the same, and do not change with doping, in the systems where doping dependence is known (see, for example, Ref. \onlinecite{reznik-rev} and \onlinecite{pint-lascuo-od}) . For such branch, the mode has a vanishing small dynamical structure factor, so we can not compare directly the dispersion close to zone center and $q$=2$\frac{\pi}{a}$. On the other hand, IR data shows well defined phonons mode at the zone center, but only in the insulating phases. We remark that the IR data assigned  to the EU(1) mode on heavily underdoped samples, corresponds very well to the frequency we found at the zone boundary for both $\Delta_1$ and $\Delta_3$ Cu-O bond stretching modes. This suggests that the physics for this mode is very similar to the one of La$_{2-x}$Sr$_x$CuO$_{4+\delta}$\cite{pint-lascuo-od,reznik-rev}, despite the present lack of dispersion data in the undoped parent compound, and of mode frequencies an the Zone Center $\Gamma$ in the doped one.  

\begin{figure}
	\includegraphics[scale=0.45,angle=0]{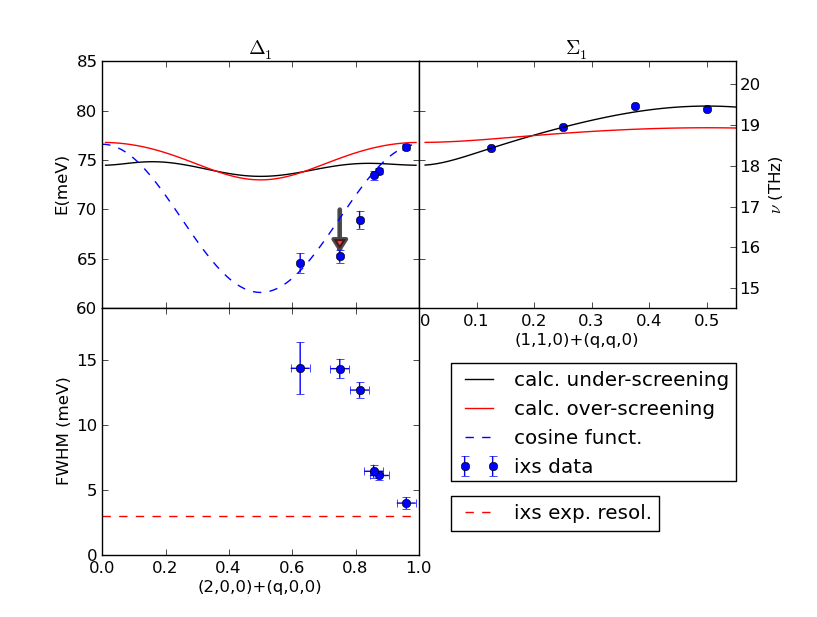}
	\caption{\label{comp-d4vsd2-w-CuO_BS_ccoc} (Color on line) Top panels: Ca$_{2-x}$CuO$_2$Cl$_2$ Cu-O bond-stretching dispersions (blue circles) measured by IXS and calculated frequencies using a shell model of lattice dynamics. Top, left: along ($q$ 0 0) ($\Delta_1$). Top, right: along ($q$ $q$ 0) ($\Sigma_1$). 
	Black lines corresponds to a model with lower doping and a screening wave number $\kappa_s = \frac{\pi}{2a}$, while red lines corresponds to a model with higher doping, and a screening wave number $\kappa_s = \frac{\pi}{a}$. The blue dashed line is a cosine function compared to the data of $\Delta_1$ symmetry in order to show the marked anomaly at  $q \sim$ 0.75. 
	Bottom panel: Blue full circles are the Full Width Half Maximum of the IXS data corresponding to the Cu-O bond-stretching mode along ($q$ 0 0), showing a maximum around $q \sim$ 0.75\cite{Matplotlib}.
	 }
\end{figure}

In Fig. \ref{comp-d4vsd2-w-CuO_BS_ccoc}, top panels, we shows the detail of the Cu-O bond stretching  for the direction with $\Delta_1$ (half-breathing) and $\Sigma_1$ (full-breathing) character and compared to our lattice dynamics simulations, with different screening wave number 
$\kappa_s$.  

In the same figure, bottom panel, we can follow the corresponding broadening of the mode, which starts at a value of less than 1 meV, once instrumental resolution subtracted, to end up at more than 12 meV.  
Note that the softening and simultaneous broadening is observed only for the mode with character $\Delta_1$, propagating along $(q 0 0)$ (Fig. \ref{comp-d4vsd2-w-CuO_BS_ccoc}, left, top and bottom). The same mode propagating along the diagonal ($\Sigma_1$, Fig. \ref{comp-d4vsd2-w-CuO_BS_ccoc}, left) is dispersing upward, with a line-shape that is resolution limited, as in other cuprates \cite{reznik-rev}.

STM experiments  in oxychlorides \cite{hanaguri} report a modulation of the  Fourier transform of the  differential conductance map $g (\mathbf{r}, E )$ at wave-vectors of ($n q$, 0 0) with $n$ integer and $q \sim$ 0.25, \textit{i.e.} for modulations with a periodicity corresponding to 4$a$. 
This is reminiscent of what happens in cuprates \cite{reznik-gunn,reznik-rev} and other perovskite-like ternary transition metal oxides \cite{weber}, with the difference that in oxychlorides, there is no observation of static diffraction by charge density modulation. We can then confirm that the observation of STM has a dynamical precursor at room temperature, as for the stripes in other cuprates. Note that the pattern observed by STM where first labelled as "checkerboard" \cite{hanaguri}, although this was not confirmed by further studies\cite{kohsaka}, that maintained, however, a Cu-O-Cu bond-centered electronic pattern without long-range order but with 4-lattice-constant-wide unidirectional electronic domains \cite{kohsaka}. 
Indeed, for a "checkerboard" pattern, we should have observed the same phonon anomaly along both the $\Delta_1$ and the $\Sigma_1$ directions, as it is the case in those transition metal oxides where such a modulation is present\cite{weber}. Here we observe the anomaly only for the mode with $\Delta_1$ character in agreement with other cuprates showing a stripe charge modulation pattern \cite{reznik-gunn,reznik-rev}. 

Regardless of the exact pattern of the charge modulation, we note that such anomaly can not be accounted for in simple models based on Landau liquid band type as DFT ones  \cite{reznik-gunn}, as the charge modulations themselves can not be simulated in such theoretical framework. 
We remark also that the propagation vector ($n$0.25, 0 0) of these unidirectional electronic domains, where we observe the  "sharp dip" anomaly (Fig.  \ref{comp-d4vsd2-w-CuO_BS_ccoc}), would corresponds to a nesting vector of the Fermi Surface, as reported for Na-doped Ca$_{2}$CuO$_2$Cl$_2$ \cite{k-shen}, although the region of the Fermi Surface connected by this vector does not show well defined \textit{quasi-particles}.

Apart from this peculiar anomaly close to the charge modulation wave-vector ($\pi/2a$ 0 0), it has been claimed that the overall bending down of the mode with  $\Delta_1$ character is well reproduced by DFT calculations \cite{bohnen,falter-lsco,giustino} that reproduce a smooth cosine function shape. These calculations found that such mode has one of the highest coupling with charge carriers. Nevertheless, the calculated electron-phonon coupling was still too low to account for the strong kink in ARPES dispersion \cite{giustino}, and the very high $T_c$ of cuprates \cite{bohnen}. 
Moreover, similar calculations showed a concurrent strong downward bending of the $\Sigma_1$ mode, that can lead to discrepancies at the zone boundary X  ($\pi/a$ $\pi/a$ 0) between calculated and experimental frequencies in optimal or slightly under-doped compounds up to 10 meV \cite{bohnen, falter-lsco}. It was suggested that such a difference originates from unrelated effects over the two directions ($q$, 0, 0) and ($q$, $q$, 0)\cite{bohnen}. 
Note that Giustino and co-workers in Ref. \onlinecite{giustino} use the label \textit{half-breathing} and \textit{full-breathing} to distinguish Cu-O bond-stretching modes with different character, as in other works\cite{pint-lascuo-od}, but the \textit{full-breathing} label is not consistent throughout literature, as it indicates modes with $\Sigma_1$ character in Ref. \onlinecite{pint-lascuo-od}, while misleadingly label a dispersion along ($q$ 0 0) with $\Delta_3$ character in Ref. \onlinecite{giustino}. 
It is worthwhile to notice that the later dispersion has been found to be flat for all dopings, either experimentally and theoretically. 
At the same time, the authors of Ref. \onlinecite{giustino} do not show their results along $\Sigma_1$, where a very bad agreement is found in other DFT calculations \cite{bohnen,falter-lsco} and experimental data. We therefore avoid the "breathing" notation.

Contrary to DFT simulations, our simple lattice dynamical calculation based on a shell model, reproduces well the dispersion of the mode with $\Sigma_1$ character, while missing the softening of the one with $\Delta_1$ character, when using a screening wave number of $\kappa_s = \frac{\pi}{2a}$ (see Fig. \ref{comp-d4vsd2-w-CuO_BS_ccoc}). When increasing the doping, and consequently the screening wave number, we observe a modest softening for both bands. The effect is relatively weak in our calculations, as we have kept all other parameters fixed, in order to allow a direct comparison between the two cases. 
The important point here is that the softening affects exclusively the modes with $\Delta_1$ and $\Sigma_1$ character, with a trend that reproduces the experimental one with increasing doping.  
This suggests that overestimated screening effects can lead to simulations that reproduce the dispersion at much higher doping.  
If compared to optimal or slightly under-doped compounds, this gives a dispersion which is accidentally similar to the one for the mode with $\Delta_1$ character, while giving a large difference between calculation and experimental dispersion for the mode with $\Sigma_1$ character, and accordingly underestimate the electron-phonon coupling. Indeed, the dispersion found in Ref. \onlinecite{bohnen,giustino} for the Cu-O bond-stretching modes with  $\Delta_1$ and $\Sigma_1$ character are strikingly similar to  the experimental dispersion found only in highly overdoped cuprates  La$_{2-x}$M$_x$CuO$_4$ (M=Sr,Ba)\cite{pint-lascuo-od,dastuto-laba}. 
Therefore, a possible unique origin for the softening of modes with both $\Delta_1$ and $\Sigma_1$ character in DFT calculation can not be ruled out. This would imply that the phonon simulations by DFT correspond to a much higher doping, that, because of the larger screening, have a lower electron-phonon coupling. To confirm such conclusion also in oxychlorides, it would be important to compare with DFT calculations, and to have data on the un-doped parent compound. 

\section{Conclusions}

We presented an extended study of the phonon dispersion in superconducting Ca$_{2-x}$CuO$_2$Cl$_2$, doped with Ca vacancies, close to optimal doping. We interpret the results using lattice dynamical calculations based on a shell model, and we compare them to what obtained for other hole-doped cuprates, in particular with the T-structure.
We found an overall agreement between our model and the data, confirming the good choice of parameters of the common interatomic potential for the lattice dynamics of cuprates first proposed in Ref. \onlinecite{chaplot}. 
An anomalous dip in the Cu-O bond stretching dispersion develops also in oxychlorides 
with a simultaneous marked broadening of the mode. The broadening maximum appears at $\approx (\pi / (2a) ~ 0 ~ 0)$ that corresponds to the charge-modulations propagation vector, observed by means of STM experiments \cite{hanaguri}.  Our data thus unveil a dynamic precursor state
of the above modulations. 
By comparing our measurements on the longitudinal Cu-O bond-stretching mode in superconducting Ca$_{2-x}$CuO$_2$Cl$_2$ with our model, we found that screening effect may bend downwards simultaneously the dispersions along both the ($q$ 0 0) and ($q$ $q$ 0) directions, a result that can help explaining the discrepancies between phonon dispersion based on DFT calculations\cite{falter-lsco,bohnen} and the experimental data in cuprates. 

\begin{acknowledgments}
We are very grateful to M. Calandra and F. Mauri for useful discussion. We also thank S. Chaplot and M. Rao for encouraging exchanges and suggestions, and to R. Heid and T. Hanaguri for enlightening discussions and critical reading of the manuscript. 
We acknowledge  D. Gambetti and B. Baptiste for technical help during measurements and A. Shukla for redaction suggestions. 
This work was supported by ESRF through Experiment No. HS-3461. I.Y. gratefully acknowledges the European Commission for financial support under the contract MIIF-CT-2006-0374.
\end{acknowledgments}

%




%
%

\end{document}